\documentclass[twocolumn,prd,showpacs]{revtex4}
\usepackage{amssymb}
\usepackage{amsmath}
\begin{document}

\title{Standing gravitational waves from domain walls}

\author{Merab Gogberashvili${}^{1,2,4}$}
\email{gogber@gmail.com}
\author{Shynaray Myrzakul${}^{3, 4}$}
\email{shinaray_81@mail.ru}
\author{Douglas Singleton ${}^{4, 5}$}
\email{dougs@csufresno.edu}

\affiliation{${\ }^{(1)}
$Andronikashvili Institute of Physics,
6 Tamarashvili Street, Tbilisi 0177, Georgia \\
${\ }^{(2)}$Javakhishvili Tbilisi State University,
3 Chavchavadze Avenue, Tbilisi 0128, Georgia \\
${\ }^{(3)}$ Department of General and Theoretical Physics,
Gumilev Eurasian National University, Astana, 010008, Kazakhstan \\
${\ }^{(4)}$ California State University, Fresno, Physics Department, Fresno,  CA 93740-8031, USA \\
${\ }^{(5)}$Institute of Gravitation and Cosmology, Peoples' Friendship University of Russia,
Moscow 117198, Russia}

\date{\today}

\begin{abstract}
We construct a plane symmetric, standing gravitational wave for a domain wall plus a massless scalar field. The scalar field can be associated with a fluid which has the properties of `stiff' matter, i.e. matter in which the speed of sound equals the speed of light. Although domain walls are observationally ruled out in the present era the solution has interesting features which might shed light on the character of exact non-linear wave solutions to Einstein's equations. Additionally this solution may act as a template for higher dimensional 'brane-world' model standing waves. 
\end{abstract}


\maketitle

%%%%%%%%%%%%%%%%%%%%%%%%%%%%%%%%%%%%%%%%%%%%%%%%%%%%%%%%%%%%%%

\section{Introduction}

Most discussions of gravitational waves involve taking the weak field limit of Einstein's field equations thus ignoring the non-linear character of the theory. There are examples of wave solutions to the full non-linear field equations such as the Einstein-Rosen cylindrical waves \cite{ew}, or non-linear gravitational plane wave solutions \cite{exact}. These exact wave solutions to the full non-linear equations are useful for understanding gravitational waves without the weak field assumption. 

In this article we present an exact standing gravitational wave solution in the presence of a domain wall and a massless scalar field. The solution we give here is a non-trivial modification of the usual domain wall solution 
\begin{equation} \label{domain-wall}
ds^2 = \frac{1}{\sqrt{1 + k |z|}}\left(dt^2-dz^2\right)
-(1 + k |z|) \left[ dx^2 + dy^2 \right]
\end{equation}
i.e. it is not simply a weak field gravitational wave in the background \eqref{domain-wall}, but is a solution 
to the full non-linear gravitational field equation. Unlike the pure domain wall solution, which is
a vacuum solution everywhere except at the location of the wall i.e. $z=0$, our solution is supported by
a scalar field. 

Vacuum standing wave solutions to the Einstein field equations were investigated in \cite{stephani} and it was found that some special case of the Einstein-Rosen cylindrical waves may be considered standing waves. However because of the non-linearity of general relativity one can not (as in electrodynamics) take two oppositely traveling plane waves (i.e. the pp-wave solutions \cite{stephani2}) and add them together to get a standing wave. Our plane symmetric, standing wave solution is supported by a domain wall and scalar field. We find that the scalar field and the transverse metric components oscillate with a $\pi /2$ phase difference which suggests a passing of energy back and forth between the scalar and gravitational field. Since the energy of the scalar field can be localized this seems to imply that one can for this solution have a localized energy for the gravitational field -- or at least for the wave part of the gravitational field. 

Defining a localized gravitational field energy is difficult \cite{energy}, and strictly speaking it is impossible since via the equivalence principle one can always locally go to a frame in which space-time is Minkowski, i.e. a frame where the gravitational field vanishes. In the conclusion we will discuss in what limited sense it may be valid to think of localized gravitational field energy for this solution. 

The requirement of a domain wall in our solution is not surprising since even in electrodynamics one must have conducting planes in order to make standing waves. In the conclusion we argue that -- minus the scalar field -- there is a large similarity between the present solution and the simple electro-magnetic standing wave between two infinite conducting planes. For the solution in this paper there is only one plane -- the domain wall at $z=0$. The other 'wall' is provided by the static part of the gravitational field which increases as one moves away from $z=0$. This is similar to the experiments of \cite{nature} where neutrons were trapped between a reflecting surface at the surface of the Earth and the gravitational potential of the Earth. As in these neutron experiments we find discrete frequencies of oscillation for our gravitational standing wave just as the neutrons had discrete frequencies/energy levels. Also this analogy between the present gravitational solution and the neutrons experiments (where the oscillating part of the gravitational field is equivalent to the neutrons) again is suggestive that at least for the wave part of the gravitational field one has some idea of localization of gravitational energy. For neutrons there is a well defined concept of localization of energy. 

%%%%%%%%%%%%%%%%%%%%%%%%%%%%%%%%%%%%%%%%%%%%%%%%%%%%%%%%%%%%%%%%%%%%%%%%

\section{Metric and field equations}

We begin by modifying the domain wall metric of \eqref{domain-wall} by
adding $t$ and $z$ dependent ansatz functions $S(t,z)$ and $V(t,z)$ as follows
\begin{eqnarray} \label{metric}
ds^2 = \frac{e^{-S(t,z)}}{\sqrt{1 + k z}}\left(dt^2-dz^2\right) - \nonumber \\
-(1 + k z) \left[ e^{V(t,z)} dx^2 + e^{-V(t,z)} dy^2 \right]~.
\end{eqnarray}
Here $k > 0$ is a constant. Note that we write $(1 + k z)$ instead of the standard domain wall function $(1 + k |z|)$ so our solution applies only in the region $z>0$. To obtain the equivalent solution for $z<0$ we could set $k<0$. If one sets $k=0$ and $S =0$ the form of the ansatz in \eqref{metric} is similar to a special case of the traveling wave ansatz considered in 
\cite{bondi, mtw}. The ansatz functions of these latter references are functions of $u=t-z$ or $v=t+z$ as one would expect for traveling waves, whereas the ansatz functions of the metric in \eqref{metric} are simply functions of $z$ and $t$ without any special form, as one would expect for standing waves. With $k \ne 0$ and $S \ne 0$ one finds that the ansatz in \eqref{metric} is some combination of the domain wall solution of \cite{taub1} and the colliding plane wave solutions of \cite{yurt}. The metric in \eqref{metric} differs from the colliding plane wave solutions in that the domain wall function, $(1 + kz)$, replaces the time-like coordinate, $t$, used in \cite{yurt}. It is important to stress the the standing wave solution we find below
is not a linearized perturbative wave solution on some background, but is an exact solution. This solution is therefore
mathematically distinct from situations such as, for example, the study of aspherical perturbed evolution of a relativistic
star interacting with gravitational waves.   

In addition to the metric ansatz given above we take a massless scalar field, $\phi (t,z)$, which obeys the Klein-Gordon equation in the background from \eqref{metric}
\begin{eqnarray} \label{phi}
\frac{1}{\sqrt{-g}}~\partial_\mu (\sqrt{-g} g^{\mu\nu}\partial_\nu
\phi) = \Box\phi = \nonumber \\
= \sqrt{1 + k z} ~ e^{S(t,z)}\left(\ddot \phi - \phi'' - \frac{k}{1+kz} \phi'\right) = 0 ~,
\end{eqnarray}
where overdots mean time derivatives and primes stand for derivatives with respect to $z$. (A system similar to that in \eqref{metric} and \eqref{phi} was recently investigated in terms of cosmological solutions \cite{dunn}).

The energy-momentum tensor of $\phi$ is given by 
\begin{equation} \label{emt-phi}
T_{\mu \nu} = \partial_\mu \phi\partial_\nu \phi - \frac 12 g_{\mu\nu} \partial^\alpha
\phi~\partial_\alpha \phi~.
\end{equation}
For this energy-momentum tensor one can write the Einstein equations,
\begin{equation}
R_{\mu \nu} - \frac{1}{2} g_{\mu \nu} R = \frac{8 \pi G}{c^4} T_{\mu \nu}~,
\end{equation} 
in the simplified form
\begin{equation}
\label{field-eqns1}
R_{\mu \nu} = 2 \partial_\mu \sigma\partial_\nu \sigma ~.
\end{equation}
Here the gravitational constant is absorbed via the redefinition of the scalar field 
\begin{equation} \label{sigma}
\sigma =\frac{\sqrt{4 \pi G}}{c^2}\phi ~. 
\end{equation}

It can be shown that the equations \eqref{field-eqns1} are equivalent to the Einstein equations for a perfect fluid, which obeys the equation of state of stiff matter. A general perfect fluid has as energy-momentum tensor given by 
\begin{equation} 
\label{fluid}
T_{\mu\nu} = (\epsilon + p) u_\mu u_\nu - g_{\mu\nu}p~,
\end{equation}
where $\epsilon$ is the energy density, $p$ is the pressure of the fluid, and $u^\nu$ is the normalized 4-velocity vector, $u^\mu u_\mu = 1$. A stiff fluid has the equation of state $\epsilon = p$ and thus the speed of sound in the fluid equals the speed of light \cite{zel}. By making to following associations: 
\begin{equation}
\label{stiff-fluid}
\epsilon  =  p = \partial_\nu \sigma
\partial^\nu \sigma~, \qquad 
u^\mu = \frac{\partial^\mu
\sigma}{\sqrt{\partial^\nu \sigma\partial_\nu \sigma}}~,
\end{equation}
it was shown that a non-rotating stiff fluid is equivalent to a massless scalar field (\ref{sigma}) \cite{taub}. 

Inserting \eqref{metric} into \eqref{field-eqns1} we obtain
\begin{eqnarray} \label{field-eqns2}
R_{tt} &=& \ddot S - S'' - \frac{k}{1+kz} S'
-\dot V^2 = 4\dot \sigma ^2~, \nonumber \\
R_{xx} &=& -\ddot V + V'' + \frac{k}{1+kz} V' =0 ~, \nonumber \\
R_{yy} &=& -\ddot V + V'' + \frac{k}{1+kz} V' =0~, \\
R_{zz} &=& -\frac{k}{1+kz} S' -V'^2 + S'' - \ddot S 
= 4 \sigma '^2~, \nonumber \\
R_{zt} &=& -\frac{k}{1+kz} \dot S - \dot V V' = 4 \dot \sigma \sigma '~. \nonumber
\end{eqnarray}
With some slight manipulation one can re-write the system of equations for $\sigma (t,z), S(t,z)$ and $V(t,z)$ as
\begin{eqnarray}
\label{field-eqns3}
\ddot\sigma - \sigma'' - \frac{k}{1+kz}\sigma ' = \ddot V - V'' - \frac{k}{1+kz} V' = 0 ~, \nonumber \\
\dot S = -\frac{1+kz}{k} \left( V' \dot V + 4\dot \sigma\sigma '\right) ~, \\
S' = - \frac{1+kz}{2k} \left(V'^2 + \dot V^2 + 4\dot \sigma^2 + 4\sigma '^2 \right) ~. \nonumber
\end{eqnarray}
Note that the scalar field, $\sigma$, and the metric ansatz function, $V(t,z)$, associated with the $x, y$ coordinates obey the same equation. In the next section we present our standing wave solution to this system. 

%%%%%%%%%%%%%%%%%%%%%%%%%%%%%%%%%%%%%%%%%%%%%%%%%%%%%%%%%%%%%%%%%%%%%%%%

\section{Standing wave solution}

One can show that 
\begin{eqnarray} \label{solution}
V(t,z) & = & A ~J_0 \left(\frac{\omega}{k} + \omega z \right) \cos (\omega t)~, \nonumber \\
\sigma (t,z) & = & \frac{A}{2} J_0 \left( \frac{\omega}{k}+\omega z \right) \sin (\omega t) ~,
\end{eqnarray}
solves the two equations from the first line of \eqref{field-eqns3}, where $J_0$ is the ordinary Bessel function of zeroth order and $A$ is the wave amplitude. In addition plugging these functions into the second line of \eqref{field-eqns3} gives $\dot S = 0$, i.e. the ansatz function $S(t,z) = S(z)$ is time-independent. This roughly corresponds to intuition from the electrodynamics, where it is only the fields and potentials in the transverse $x,y$ directions which are time dependent. Finally taking $V(t,z)$ and $\sigma (t,z)$ from \eqref{solution}, inserting them into the last equation of \eqref{field-eqns3} and integrating gives
\begin{eqnarray} \label{solution1}
S(z)  =  C -\frac{A \omega ^2}{4k^2} (1 + kz)^2 \left[ J_0^2\left( \frac{\omega}{k}+\omega z \right) + \right. \nonumber \\
\left. + ~2 J_1^2 \left( \frac{\omega}{k}+\omega z \right) -  J_0 \left( \frac{\omega}{k}+\omega z \right)J_2 \left( \frac{\omega}{k}+\omega z \right) \right] ~.
\end{eqnarray} 
Here $C$ is an integration constant and $J_1$ and $J_2$ are ordinary Bessel functions of first and second order respectively. 

In order to interpret this solution as a domain wall embedded in some scalar field which extends to the right hand side (i.e. $z>0$) we should recover the pure domain wall metric as $z \rightarrow 0$ which is the location of the delta function energy-momentum tensor source of the domain wall. In other words we want $V(t,z) \rightarrow 0$ and $S(z) \rightarrow 0$ as $z \rightarrow 0$. In the limit $z \rightarrow 0$ we find for $V(t,z)$
\begin{equation} \label{bc-v}
V(t,0) \rightarrow A~J_0 \left(\frac{\omega}{k} \right) \cos (\omega t) ~.
\end{equation}
To have $V(t,0) \rightarrow 0$ we need to chose $\omega / k = z_n$ where $z_n$ is the $n^{th}$ zero of the $J_0$ Bessel function. If we take the properties of the domain wall as fixed (i.e. $k$ or the domain wall energy density/tension is fixed) then the relationship implies that only certain discrete frequencies of oscillation are allowed, namely $\omega = k z_n$. Note also that the scalar field, $\sigma$, will also vanish at $z=0$ under the condition $\omega = k z_n$. This condition also makes the first and last term in brackets of \eqref{solution1} vanish. To cancel the middle term one needs to choose the integration constant as
\begin{equation} \label{int-const}
C = \frac{A \omega ^2}{2 k^2}  J_1^2 (z_n) ~,
\end{equation}
where again $z_n = \omega /k$.

Examining the energy density associated with the scalar field one can show that its energy density is positive definite. Using the expression for $\sigma$ from \eqref{solution} in \eqref{emt-phi} we find
\begin{eqnarray}
T_{00} = 2 \partial_0 \sigma \partial_0 \sigma - g_{00} \left (\partial^0 \sigma \partial_0 \sigma
+ \partial^z \sigma \partial_z \sigma \right) = \nonumber \\
= \dot \sigma^2 + \sigma '^2 = \frac{A^2 \omega ^2}{4} \left[
J_0^2 \left( \frac{\omega}{k}+\omega z \right) \cos^2 (\omega t) + \right. \\
\left. + J_1^2\left( \frac{\omega}{k}+\omega z \right) \sin ^2(\omega t) \right] \ge 0~. \nonumber
\end{eqnarray}
However, the stiff fluid version of the matter source (i.e. \eqref{fluid} plus
\eqref{stiff-fluid}) has an energy density which oscillates between positive and negative values. This can be seen by inserting $\sigma$ from \eqref{solution} into the stiff fluid energy density $\epsilon = \partial_\nu \sigma \partial^\nu \sigma$. Currently many types of matter are considered which violate some or all of the energy conditions that were formerly imposed on matter sources in general relativity (e.g. dark energy \cite{riess}, phantom energy \cite{caldwell}). Therefore one might chose not to worry about the non-positive definite feature of the stiff fluid. Alternatively one could introduce a second fluid which acts as a cosmological constant and alter the stiff fluid to an almost stiff fluid as in \cite{merab} where it was shown that this gave a positive definite energy density for the fluid version of the matter source. 

The standing wave solution given by \eqref{metric} and \eqref{solution} can be said to have a single polarization. For traveling waves one can write a linearized form of the metric \eqref{metric} as (see section 15.2 of \cite{stephani2}): 
\begin{eqnarray}
\label{linear-wave}
ds^2 &=& dt^2 - \left(1+f_{xx}\right) dx^2 - 2 f_{xy} dx~dy - \nonumber \\
&-& \left(1+f_{yy}\right) dy^2  - dz^2 ~, 
\end{eqnarray}
where $f_{xx, yy, xy} = a_{xx,yy,xy} F(u)$ and $u=t-z$; $F(u)$ can be any function, although usually it is taken to be some sinusoidal function, e.g. $F(u) = \cos(u)$ or $\sin (u)$. 

There are two different 'linear' polarizations that are possible \cite{stephani2, mtw}: 
(i) $a_{xx} = -a_{yy}$ and  $a_{xy}=0$, or (ii) $a_{xy} = a_{yx} \ne 0$ and $a_{xx}=a_{yy}=0$. 
In the limit of large $z$ using the standard expansion of Bessel functions of the first kind
\begin{equation}
\label{bessel}
J_n (\omega z) \approx \sqrt{\frac{2}{\pi \omega z}} \cos \left[ \omega z
-\left( n +\frac{1}{2} \right) \frac{\pi}{2} \right] ~,
\end{equation}
the ansatz function $V(z,t)$ becomes:
\begin{eqnarray}
\label{v-approx}
V(z,t) &=& A J_0 \left(\frac{\omega}{k} + \omega z \right) \cos (\omega t) \approx \nonumber \\
&\approx& A \sqrt{\frac{2}{\pi \omega z}} \cos \left( \omega z- \frac{\pi}{4} \right) \cos(\omega t) ~.
\end{eqnarray}
Then expanding $\exp[\pm V(z,t)] \approx 1 \pm V(z,t)$ the $dx^2, dy^2$ part of the metric
\eqref{metric} becomes:
\begin{equation}
\label{metric-approx}
-kz \left[ (1+ V) dx^2 + (1-V) dy^2 \right] ~,
\end{equation}
where $V(z, t)$ is given by \eqref{v-approx}. Ignoring the overall pre-factor of $kz$, and comparing \eqref{metric-approx} with the $dx^2, dy^2$ part of the metric \eqref{linear-wave}, we find that the solution given by \eqref{metric} with \eqref{solution} has linear polarization,
\begin{equation}
a_{xx} = A \sqrt{\frac{2}{\pi}} = -a_{yy}~, ~~~~~ a_{xy}=0~,
\end{equation} 
corresponding to case (i) for the metric \eqref{linear-wave}.

%%%%%%%%%%%%%%%%%%%%%%%%%%%%%%%%%%%%%%%%%%%%%%%%%%%%%%%%%%%%%%%%

\section{Discussion of physical meaning of solution}

In this article we have given a simple example of a standing gravitational wave with plane symmetry. This solution is not a vacuum solution but requires the presence of a domain wall at $z=0$ and a massless scalar field, $\sigma$, occupying the region $z>0$. From \eqref{solution} one finds that the metric function, $V(t,z)$, and the scalar field, $\sigma (t,z)$, had the same spatial dependence but their time oscillations where $\pi /2$ out of phase. One could view this as the energy of the oscillation passing back and forth between the scalar field and the wave (time-dependent) part of gravitational field, as embodied in the metric ansatz function $V(z,t)$. Since the field energy of the scalar field can be localized this phase connection between the oscillations of the scalar field and the gravitational field suggests that for this solution one might be able to define a local gravitational field energy -- at least for the wave (time-dependent) part of the solution. As mentioned in the introduction a definition of local gravitational field energy is strictly impossible because of the equivalence principle. Nevertheless the phase relationship between the scalar field and the $dx^2$, $dy^2$ components of the metric ($\exp[\pm V(z,t)]$ factors) is suggestive of the wave part of the gravitational field carrying a localized energy. Note, since this solution is not asymptotically flat and the domain wall is not a localized source, one can not define global gravitational field energy via surface integrals over effective energy-momentum tensors as discussed in section 14.3 of reference \cite{stephani2} or section 20.5 of reference \cite{mtw}.

Absent the matter (i.e. the scalar field) needed to support this gravitational standing wave, there is a lot in common between the present solution and the simple electromagnetic standing wave between two infinite conducting planes. One plane is obviously the domain wall at $z=0$ but where is the other plane? A possible answer lies in the static, Newtonian potential 
\begin{equation} \label{Phi}
\Phi (z) = \frac 12 \left[g_{00}(z) - g_{00}(z=0)\right]~.
\end{equation}
In the limit of large $z$ the ansatz function $S(z)$ from \eqref{solution1} takes the form
\begin{eqnarray}
\label{newton}
S &\approx& -\frac{A \omega z}{2 \pi} \left[ \cos ^2 \left( \omega z - \frac{\pi}{4} \right)
+2 \cos ^2 \left(\omega z - \frac{3 \pi}{4} \right)\right. -\nonumber \\
&-& \left. \cos \left(\omega z - \frac{\pi}{4} \right)\cos \left(\omega z - \frac{5 \pi}{4} \right) \right] = \nonumber \\
&=& -\frac{A \omega z}{\pi} ~.
\end{eqnarray}
In the first step we have used \eqref{bessel} and in the last step we have combined all the cosines using standard identities. Using these results the Newton potential \eqref{Phi} takes the asymptotic form
\begin{equation}
\label{newton2}
\Phi (z) = \frac{\exp[-S(z)]}{2\sqrt{1 + kz}} -\frac{1}{2} \approx
\frac{\exp[A \omega z / \pi ]}{\sqrt{kz}} 
\end{equation}
This exponentially increasing, static potential may be thought of as trapping the orthogonal oscillations of the gravitational field (i.e. those associated with the metric components $g_{xx}, g_{yy}$ and the ansatz function $V(z, t)$). In other words it is the increasing static gravitational field associated with $g_{00}$, which traps the oscillatory parts of the gravitational field associated with $g_{xx}, g_{yy}$, thus acting as the second (soft) plane in conjunction with the (hard) plane of the domain wall at $z=0$. Also note that just as electromagnetic standing waves have discrete frequencies, so too the plane gravitational standing waves have discrete frequencies. For the the present system one finds $\omega = k z_n$, where $z_n$ are the zeros of the $J_0$ Bessel function and $k$ is the energy density/tension of the domain wall. 

The solution presented here also has similarities to the neutron trapped in a gravitational field experiments of \cite{nature}. In these experiments cold neutrons were trapped between a reflecting surface on the surface of the Earth (the domain wall in this paper) and at their upper extent by the Earth's gravitational field (the $z \rightarrow \infty$ gravitational field of the domain wall and scalar field in this paper). The energy of the neutrons in these experiments was quantized as is the case of the gravitational standing waves of the present paper. Note that for neutrons there is a well defined concept of local energy. 

One can combine the various asymptotic forms of the ansatz functions given in \eqref{v-approx} and \eqref{newton} to write the metric \eqref{metric} as:
\begin{eqnarray}
\label{metric-approx2}
ds^2 &\approx& \frac{\exp[A \omega z / \pi ]}{\sqrt{kz}} ( dt^2 -dz^2 ) -\\
&-&kz\left[1+A \sqrt{\frac{2}{\pi \omega z}} \cos \left( \omega z- \frac{\pi}{4} \right) \cos(\omega t) \right] dx^2 -
\nonumber \\
&-&kz\left[1-A \sqrt{\frac{2}{\pi \omega z}} \cos \left( \omega z- \frac{\pi}{4} \right) \cos(\omega t) \right] dy^2~.
\nonumber
\end{eqnarray}
Written in this asymptotic form one sees that the metric splits into the usual, time-dependent 
standing wave form (i.e. the  $\pm A \sqrt{\frac{2}{\pi \omega z}} \cos 
\left( \omega z- \frac{\pi}{4} \right) \cos(\omega t)$
terms) sitting in a static background (i.e. the $\frac{\exp[A \omega z / \pi ]}{\sqrt{kz}}$
terms). This split gives further support to the suggestion that for this solution the phase relation between the 
scalar field and ansatz function, $V(z,t)$, may allow one to think about a local gravitational field energy for the 
time-dependent oscillating part of the metric -- but not the static, background part of the metric.

Although this system is not realistic in the sense that domains walls are ruled out (in the present era) by observational evidence \cite{okun} it is nevertheless an example of an exact, plane standing wave solution to the full non-linear Einstein field equations, which has many similarities to the simple electro-magnetic plane standing wave. Furthermore there are two features of this solution which may point to some kind of localization of gravitational field energy: (i) the oscillation with $\pi / 2$ phase difference between the scalar field, $\sigma$, and the transverse components of the metric, $g_{xx}, g_{yy}$; (ii) the apparent trapping of the transverse oscillations of the metric components $g_{xx}, g_{yy}$ by the static gravitational field associated with $g_{00}$. 

One could also use the present 4D solution to study similar standing wave solutions in 5D or higher brane world models proposed in \cite{gog}. Investigations in this direction are underway \cite{doug-merab}.

\medskip
\noindent {\bf Acknowledgments:}

M. G. and D. S. are supported by a 2008-2009 Fulbright Scholars Grants. D. S. also would like to thank Vitaly Melnikov for the invitation to work at the Institute of Gravitation and Cosmology at Peoples' Friendship University of Russia. 

%%%%%%%%%%%%%%%%%%%%%%%%%%%%%%%%%%%%%%%%%%%%%%%%%%%%%%%%%%%%%%

\end{document}